\documentstyle[aps,prd,epsf,twocolumn]{revtex} 

\draft

\begin{document}
\twocolumn[\hsize\textwidth\columnwidth\hsize\csname
@twocolumnfalse\endcsname

\title{Probing neutrino decays with the cosmic microwave background}

\author{Steen Hannestad}

\address{Institute of Physics and Astronomy,
University of Aarhus,
DK-8000 \AA rhus C, Denmark}

\date{\today}

\maketitle

\begin{abstract}
We investigate in detail the possibility of constraining neutrino decays
with data from the cosmic microwave background radiation (CMBR). Two
generic decays are considered $\nu_H \to \nu_L \phi$ and
$\nu_H \to \nu_L \bar{\nu}_L \nu_L$. We have solved the momentum dependent
Boltzmann equation in order to account for possible relativistic decays.
Doing this we estimate that any neutrino with mass $m \gtrsim 1$ eV
decaying before the present should be detectable with future CMBR data.
Combining this result with other results on stable neutrinos, any neutrino
mass of the order 1 eV should be detectable.
\end{abstract}

\vskip1.8pc]

\section{introduction}
The possibility that one or more neutrino species are unstable on a
cosmologically interesting timescale has been considered very extensively
in the literature (see for instance
\cite{BS98,WGS95,terasawa,kands,sandturn,madsen,gyuk,kawasaki,kawasaki2,hannestad1,DHP98} and references therein).
In particular it has been pointed out that cosmology
is an excellent laboratory for neutrino physics and that many exotic 
neutrino models that are inaccessible to tests in terrestrial experiments
may be tested using cosmology.

Big Bang nucleosynthesis has turned out to be a very powerful probe for this
purpose
\cite{terasawa,kands,sandturn,madsen,gyuk,kawasaki,kawasaki2,hannestad1,DHP98}.
However, one shortcoming of BBN in this regard is that it is
only sensitive to physics before the end of nucleosynthesis, at $t \simeq
10^3$ s. After this the relative abundances of light elements is fixed
until the present.

Another intriguing possibility is to use the CMBR to probe exotic neutrino
physics \cite{lopez1,lopez2,hannestad2,HR99}.
This has been done in the past to constrain neutrino decays
and other effects of non-standard neutrinos. However, there has been no
real coherent treatment of what limits can be put on neutrino decays
from CMBR data. In the present work we shall try to discuss all the
possible decay modes of a heavy neutrino in the context of CMBR.

\section{neutrino decays}

A massive neutrino can in principle have many different decay modes. The
simplest possibilities are the standard model decays 
$\nu_H \to \nu_L \gamma$ and $\nu_H \to e^+ e^- \nu_e$
\cite{moha,kimpevs}. 
However, they both suffer from the fact that their decay products contain
electromagnetically interacting particles and they can therefore be
observationally constrained.
Raffelt \cite{Raff96} provides 
an excellent review on the observational limits on such decays.
The main conclusion is that they are excluded unless the lifetime
is exceedingly long.

We are then left with other possibilities which demand more exotic models.    
We shall look at two decay modes which cover the viable possibilities
$\nu_H \to \nu_L \phi$ and $\nu_H \to \nu_L \bar{\nu}_L \nu_L$
\cite{moha,kimpevs}.
Below we shall discuss the possible scenarios where such decays can take
place.

\subsection{$\nu_H \to \nu_L \phi$}

The primary model for this type of decay is the majoron model
\cite{moha}. This is
a specific model for the generation of neutrino mass. In this model, the
neutrino is a Majorana particle and $\phi$ is a scalar or pseudoscalar.
If we assume a Yukawa type interaction of the form
\begin{equation}
{\cal L} = g \phi \bar{\nu}_L \gamma_5 \nu_H
\end{equation}
and that the daughter particles are massless we obtain a rest-frame decay
rate of
\begin{equation}
\Gamma = \frac{g^2 m_H}{8 \pi}.
\end{equation}
It is possible to constrain $g$ from data on neutrinoless double beta
decays to be $g \lesssim 10^{-4}$ \cite{kimpevs}. 
Inserting numbers we get a lower bound to
the lifetime of 
\begin{equation}
\tau \gtrsim 1.6 \times 10^{-12} \, m^{-1}_{\rm MeV} \, {\rm s}.
\end{equation}
This bound is obviously not very restrictive and by far the best bounds
on such decays come from astrophysical considerations.

\subsection{$\nu_H \to \nu_L \bar{\nu}_L \nu_L$}

This decay is somewhat more complicated.
However, three-body decays are an
equally interesting possibility which arises in left-right symmetric
models and in general in models with neutral current flavour violation.
A neutrino decay will then be mediated by either a Higgs, $\Delta_0$,
or a $Z'$ \cite{moha}. In either case, it is safe to assume that the boson
exchanged is extremely heavy so that the decay is effectively a 
four-point interaction.

An interesting question is how fast such three-body decays can proceed
in realistic models.
If one assumes a non-derivative coupling as would be
the case for both the above possibilities then one arrives at an
amplitude of
\begin{equation}
M = \frac{G}{\sqrt{2}} \bar{u}_2 O u_1 \bar{u}_4 O v_3,
\end{equation}
where the coupling $O$ is given by
\begin{equation}
O = \{I,\gamma_5, \gamma_\mu, \gamma_\mu \gamma_5\},
\end{equation}
or a combination of these operators.

Since all of the final-state particles are identical
(assuming that there is $CP$ symmetry), all of these different possibilities
give the same decay kinematics because of the indistinguishability of the
final state particles, the only difference being the absolute decay rate.
This of course simplifies the calculations a lot. We need do only 
one calculation for each value of the rest frame decay rate, $\Gamma$. This
quantity is directly related to the coupling strength $G$ by
\begin{equation}
\frac{1}{\tau} = \Gamma = C G^2 \frac{m^5}{\pi^3},
\end{equation}
where $C$ is a constant which depends on the specific form of
the operator $O$.
In the remainder of the paper we shall refer to the matrix element for this
process as 
a ``weak'' matrix element because it arises from interactions very
similar in structure to the standard weak interactions.

For comparison we also calculate how three-body decays proceed if the
decay matrix element is constant
\begin{equation}
\sum |M|^2 = K = G^2 m^4.
\end{equation}
This gives a rest frame-decay lifetime of
\begin{equation}
\frac{1}{\tau} = \Gamma = \frac{m^5}{512 \pi^3} G^2.
\end{equation}
To see what lifetimes can be expected in such a model we take an example
where the decay is through a flavour violating neutral current.
We can rewrite the coupling constant as $G^2 = \theta^2 G_F^2$ and from
this we get
\begin{equation}
\tau = 2.88 \times 10^{14} \, \theta^2 m_{\rm MeV}^{-5} \, {\rm s}.
\end{equation}
The limit on $\theta$ is of the order $\theta \lesssim 10^{-5}$ 
\cite{moha} which
means that for $m \lesssim 0.2$ MeV, the decay lifetime is longer than
the Hubble time.

Of course one might postulate a fast invisible decay mode to a three
neutrino final state, and its effect on cosmology would then be one way of
detecting it. For this reason alone it is interesting to calculate 
thoroughly how such a decay would affect cosmology.
In the following we shall just take a heuristic approach and examine
the consequences of a fast three-body decay without looking at the
theoretical background for this.


\section{formalism}

In the following we will go through the formalism needed to calculate
decay rates and distributions for both two- and three-body decays.
In general we need to solve the Boltzmann equation which has the form
\cite{bernstein}
\begin{equation}
L[f] = C_D[f],
\end{equation}
where
\begin{equation}
L[f] = \frac{\partial f}{\partial t} - H p \frac{\partial f}{\partial p}.
\end{equation}
The right hand side of the equation describes decay terms of various
sort. These are quite different for two and three body decays and we shall
discuss each in turn.

\subsection{two body decays}

Two body decays have quite simple kinematics since the rest-frame energy
of each daughter particle is well defined and equal to $m_H/2$.
We get for the decay terms \cite{starkman}

\begin{equation}
C_{\mbox{\scriptsize{dec}}}[f_{\nu_H}] = 
- \frac{m_{\nu_H}^{2}}{\tau m_{0} E_{\nu_H}
p_{\nu_H}}
\int_{E_{\phi}^{-}}^{{E_{\phi}^{+}}}dE_{\phi}
\Lambda(f_{\nu_H},f_{\nu_L},f_{\phi})
\end{equation}

\begin{equation}
C_{\mbox{\scriptsize{dec}}}[f_{\nu_L}] = 
\frac{g_{\nu_H}}{g_{\nu_L}} \frac{m_{\nu_H}^{2}}{\tau m_{0} E_{\nu_L} 
p_{\nu_L}}
\int_{E_{\nu_H}^{-}}^{{E_{\nu_H}^{+}}}dE_{\nu_H}
\Lambda(f_{\nu_H},f_{\nu_L},f_{\phi})
\end{equation}

\begin{equation}
C_{\mbox{\scriptsize{dec}}}[f_{\phi}] = 
\frac{g_{\nu_H}}{g_{\phi}}
\frac{m_{\nu_H}^{2}}{\tau m_{0} E_{\phi} 
p_{\phi}}
\int_{E_{\nu_H}^{-}}^{{E_{\nu_H}^{+}}}dE_{\nu_H}
\Lambda(f_{\nu_H},f_{\nu_L},f_{\phi}),
\label{bosdec}
\end{equation}
where
$\Lambda(f_{\nu_H},f_{\nu_L},f_{\phi}) = f_{\nu_H}(1-f_{\nu_L})(1+f_{\phi})-
f_{\nu_L}f_{\phi}(1-f_{\nu_H})$, 
$m_{0}^{2} = m_{\nu_H}^{2}-2(m_{\phi}^{2}+m_{\nu_L}^{2})+
(m_{\phi}^{2}-m_{\nu_L}^{2})^{2}/m_{\nu_H}^{2}$.
$\tau$ is the lifetime of the heavy neutrino and $g$ is the
statistical weight of a given particle. We use $g_{\nu_H} = g_{\nu_L}
 = 2$ and
$g_{\phi} = 1$, corresponding to $\phi = \overline{\phi}$ 
The integration limits are
\begin{eqnarray}
E_{\nu_H}^{\pm} (E_{i}) & = & \frac{m_{0}m_{\nu_H}}
{2m_{i}^{2}}[E_{i}(1+4(m_{i}/m_{0})^{2})^{1/2} \pm \\ \nonumber
& & (E_{i}^{2}-m_{i}^{2})^{1/2}]
\end{eqnarray}
and
\begin{equation}
E_{i}^{\pm} (E_{\nu_H}) = \frac{m_{0}}{2m_{\nu_H}}
[E_{\nu_H} (1+4(m_{i}/m_{0})^{2})^{1/2} \pm p_{\nu_H}]
\label{energyi}
\end{equation}
where the index $i = F,\phi$.

\subsection{three body decays}

The decay term in the Boltzmann equation can in this case be written as
\cite{bernstein,KT90}
\begin{eqnarray}
C_{\text{D}}[f] & = & \frac{1}{2E_{1}}\int d^{3}\tilde{p}_{2}
d^{3}\tilde{p}_{3}d^{3}\tilde{p}_{4}
\Lambda(f_1,f_2,f_3,f_4)
\label{integral}\\ 
& & \,\,\, \times \sum |M|^2 \delta^{4}
({\it p}_{1}+{\it p}_{2}-{\it p}_{3}-{\it p}_{4})(2\pi)^{4}, 
\nonumber
\end{eqnarray}
with $d^{3}\tilde{p}_{i} = d^3p_i/(2E_i(2\pi)^3)$ and
$\Lambda(f_1,f_2,f_3,f_4) = 
(f_2 f_3 f_4 (1-f_1)-f_1(1-f_2)(1-f_3)(1-f_4))$.
Details of how to evaluate this phase-space integral can for instance be 
found in Ref.~\cite{hannestad3}.

The simplest frame to evaluate the phase space integral in is the rest
frame of the parent particle. 
Since the three body decay scheme does not yield a specific energy for
the daughter particles, it is interesting to study how the energy 
distribution of daughter particles differ for the different interaction
matrix elements.
In order to do this it is very convenient
to calculate the differential production rate of daughter particles
\begin{equation}
\frac{d\Gamma}{dp} = \Gamma_0 \times \cases{8 p/m & constant,\cr 
64(p/m)^2 (1-\frac{5}{3}p/m) & ``weak''.}
\end{equation}
It should be noticed here that the maximum attainable momentum for any
daughter particle is $m/2$ in order to have 4-momentum conservation.
In Fig.~1 we have plotted the decay distributions for both cases.
The most interesting feature is that the weak matrix element leads to a
decay distribution which is stronger peaked in momentum space.

\begin{figure}[h]
\begin{center}
\epsfysize=7truecm\epsfbox{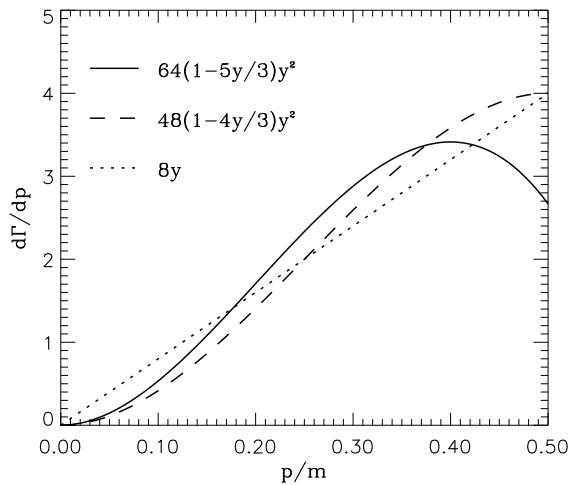}
\vspace{0truecm}
\end{center}
\baselineskip 17pt
\caption{The differential energy distribution of the light neutrino
for three different possibilities. The full line is from the ``weak''
matrix element, the dotted for the constant matrix element and the
dashed for the standard model decay $\nu_H \to \nu_L e^+ e^-$. The
lines have been normalised to $\Gamma=1$.}
\label{fig1}
\end{figure}

In the standard model decay $\nu_H \to \nu_L e^+ e^-$ one gets instead
the ``standard'' result for the decay distribution of the neutrino
\cite{gyuk}
\begin{equation}
\frac{d\Gamma}{dp} = \Gamma_0 \times 48 (p/m)^2 (1-\frac{4}{3}p/m).
\end{equation}
The reason why our result is different is that we assume that all three
final state particles are identical.
As for the average energy of the emitted neutrinos, it is of course 
equal to $m/3$ both for the both types of matrix elements. For the
standard model decay above, it is equal to $7m/20$, which is slightly
higher.

\subsection{background cosmology}

In order to get the decay into a proper cosmological setting it is then
necessary to prescribe how the universe in general evolves with time.
In the simple case where the universe is radiation dominated at decay
and the energy densities of both parent and daughters is subdominant,
the time-temperature relation is simply \cite{KT90}
\begin{equation}
t(s) = 1.71 \, g_*^{-1/2} T_{\rm MeV}^{-2},
\label{tTsimple}
\end{equation}
where $g_*$ is the effective number of degrees of freedom.
However, in general this is not true because the decay might take place
while the universe is matter dominated or partly matter dominated.
In this more general case it is necessary to use the equation of
energy conservation
\begin{equation}
\frac{d}{dt}(\rho R^{3}) + p \frac{d}{dt}(R^{3}) = 0,
\label{eq:energy}
\end{equation}
which is an exact equation, together with the Friedmann equation
\begin{equation}
H^2 = \frac{8 \pi G \rho}{3}-\frac{k}{R^2},
\label{eq:friedmann}
\end{equation}
in order to get a relation between time and temperature in the universe
\cite{KT90}.


\section{numerical solution of the Boltzmann equation}

In order to investigate the impact on the CMBR of decays it is necessary 
solve the Boltzmann equation numerically. 
To do this we have used a grid in comoving momentum space.

However, first we have to determine the initial conditions
of the system. Primarily this means that we must discuss whether or
not the daughter particles have thermal distributions prior to decay.
For the $\phi$-particle we shall assume that it is not present prior 
to decay. This will be the case if the particle decoupled from thermal
equilibrium prior to the QCD phase-transition \cite{KT90}.
For the light neutrino there are two possibilities, either $\nu_L$ is one
of the three active species or it is a sterile species. In the former
case we assume a thermal distribution prior to decay and in the
latter we assume that there are no light neutrinos present before the
decay commences.
We shall always assume that the parent neutrino has a thermal
population prior to decay.

For a qualitative discussion of the decays
it is convenient to introduce a relativity
parameter $\alpha$ for the decay \cite{hannestad1}. 
Here, we assume that the universe is
radiation dominated during the decay.
A species becomes non-relativistic roughly when $T=m/3$. Using this the
relativity parameter can be written as
\begin{equation}
\alpha \equiv 6.5 \times 10^{-2} g_*^{1/2} \tau_{\rm s} m_{\rm MeV}^2.
\end{equation}
The decay is then relativistic if $\alpha \lesssim 1$ and non-relativistic
if $\alpha \gtrsim 1$.

Here we have used the simple
relation, Eq.~(\ref{tTsimple}). 
This has been done in order to make the decay physics
more transparent, but in the next section where actual mass-lifetime limits
are calculated we use the energy conservation
and Friedmann equations, Eqs.~(\ref{eq:energy}-\ref{eq:friedmann}).
\begin{figure}[h]
\begin{center}
\epsfysize=7truecm\epsfbox{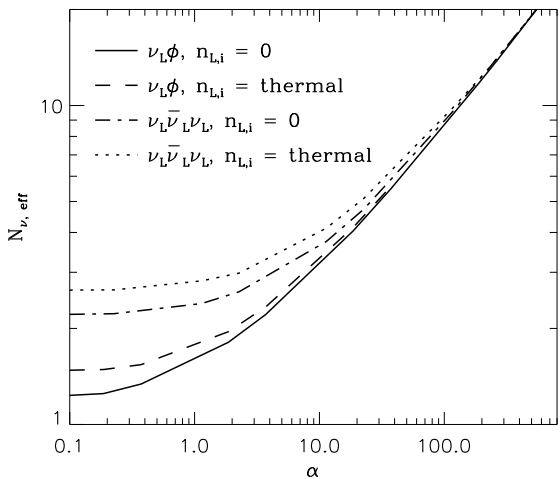}
\vspace{0truecm}
\end{center}
\baselineskip 17pt
\caption{The final energy density in decay products for the different
possible decays, parametrised in equivalent neutrino species. The standard
model with no decay would correspond to $N_{\nu, {\rm eff}}=1$.}
\label{fig2}
\end{figure}

\subsection{two body decays}

The two-body decay is interesting in that one of the final state 
particles is a boson. This means that because of stimulated emission
effects there will be a large population of low momentum $\phi$'s. 
This of course only happens if these states are accessible, i.e. for
relativistic decays \cite{madsen,starkman}
In the Appendix we have calculated the final energy density in decay
products for both very relativistic and completely non-relativistic
decays. These results are compared with the full Boltzmann solution
in Fig.~2. The full solution clearly has the correct asymptotic 
behaviour.

\subsection{three body decays}

For non-relativistic decays, the cosmic frame is equal 
to the rest-frame of the parent particle. 
In this case, for equal total decay rates, the ``weak''
and the constant matrix element give somewhat different decay
distributions.
The reason is that energy and number conservation only yields two
fixed parameters. For an equilibrium distribution this is sufficient to
describe the distribution fully by a temperature and a pseudo-chemical
potential, but the distribution of the daughter particle is not an
equilibrium one so that there are no two parameters fully describing the
distribution function.

For extremely relativistic decays the decay and inverse decay installs
equilibrium in the particle distributions for both parent and daughter.
In this case, the decay proceeds in complete equilibrium and the final
distribution functions become identical, independent of the actual
matrix element.

We have plotted the final momentum distribution of the daughter particles
for different values of the relativity parameter, $\alpha$.
Indeed one sees that for very relativistic decays the distributions become
identical. For non-relativistic decays they 
become very significantly different, even though the zeroth and first
moments are identical.
If one is interested in applications where the actual shape of the 
distribution is important, as would be the case if the daughter is an
electron neutrino present during nucleosynthesis 
\cite{hannestad1,DHP98}
or an eV particle
constituting the Hot Dark Matter \cite{hannestad4}, 
this difference in distribution could
potentially be important. However, if one is interested only in the
final number and energy density, the difference between using different
matrix elements is indeed very small and can be safely ignored.

It is quite interesting to compare the final decay distributions with
equilibrium distributions of the form
\begin{equation}
f_{\rm eq}(p) = \frac{1}{\exp\left(\frac{p-\mu}{T}\right)+1},
\end{equation}
with the same number and energy densities as the true distributions.
We have plotted these equilibrium distributions together with the full
distributions calculated from the Boltzmann equation in Fig.~3. It is
seen that in the strongly relativistic case the decay proceeds in 
equilibrium and the final distributions are of equilibrium form.
For strongly non-relativistic decays the equilibrium distributions are quite
poor approximations to the true distribution functions, as expected.

\begin{figure}[h]
\begin{center}
\epsfysize=7truecm\epsfbox{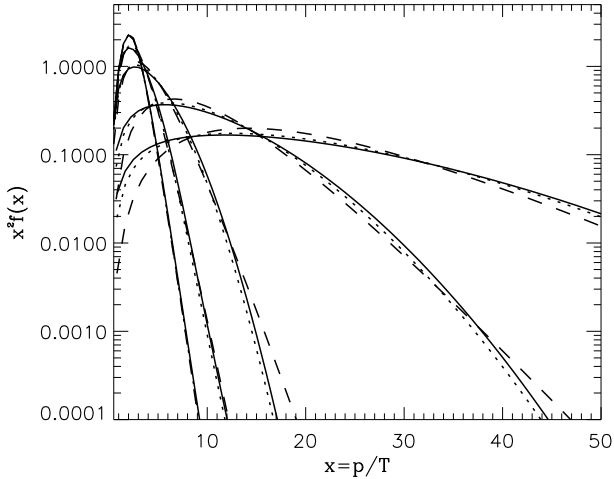}
\vspace{0truecm}
\end{center}
\baselineskip 17pt
\caption{The final momentum distribution of the light neutrino
after decay for the three body decay $\nu_H \to \nu_L \bar{\nu}_L \nu_L$.
The full line is for the ``weak'' matrix element, the dotted for the 
constant matrix element and the dashed line shows an equilibrium distribution
with the same number and energy density. The five sets of curves 
correspond from left to right to $\alpha$ = 0.2, 2, 10, 100, 500.}
\label{fig3}
\end{figure}


\section{CMBR effects}

Using the results on the energy density evolution of the decaying particle
and its daughters we can now proceed to investigate the CMBR effects
from neutrino decays.
The fluctuations are usually described in terms of spherical harmonics
\cite{coles+lucchin}
\begin{equation}
T(\theta,\phi) =\sum_{lm}a_{lm}Y_{lm}(\theta,\phi),
\end{equation}
where the coefficients are related to $C_l$ coefficients by
\begin{equation}
C_l \equiv \langle|a_{lm}|^2\rangle.
\end{equation}
These fluctuations were first detected in 1992 by the COBE satellite
\cite{cobe},
but only for $l \lesssim 20$. At such low $l$ the power spectrum is
almost degenerate in the cosmological parameters and no real constraints
are obtainable. In the next few years, however, the power spectrum
will hopefully be measured out to $l \simeq 2500$ by two new probes, 
MAP and PLANCK \cite{MAP+PLANCK}, and
using this data should yield precision measurements of the
physical parameters at recombination.
We shall here assume that MAP will deliver data out to $l=1000$ and
PLANCK to $l=2500$.

Since we do not yet have the precision data in hand it is hard to
foretell how accurately the different parameters can actually be
measured.
The usual procedure is to use what is called error forecasting
\cite{tegmark}.
This method assumes an underlying cosmological model to be the ``true''
model. From that one can calculate how sensitive the data are to changes
in the cosmological parameters.
It should be noted here that this method can at best give an estimate
of the obtainable precision. Once the data becomes available the actual
problem of determining the cosmological parameters will be much tougher 
because the whole parameter space has to be investigated.

In the present paper we shall assume as the reference the standard
cold dark matter model which can be described by a vector of values in
the many-dimensional parameter space \cite{tegmark}
\begin{equation}
\Theta = (\Omega, \Omega_b, \Lambda, h,n,\tau).
\end{equation}
As free parameters we have chosen the total density, $\Omega$,
the density in baryons, $\Omega_b$, the cosmological constant, $\Lambda$,
the Hubble parameter, $h$, the spectral index, $n$, and the optical depth
to reionisation, $\tau$. 
In addition to this we have included as free parameters the
energy density injected by neutrino decays as well as the time at which
the energy is injected.

Our choice of reference model is ``standard'' cold dark matter
\begin{equation}
\Theta_{\rm CDM} = (1, 0.08, 0, 0.5,1,0).
\end{equation}
Note that we have chosen a limited set of free parameters. In principle
one should use a parameter space consisting of all possible free
parameters.

Our procedure will then be the following: for given lifetime $\tau_H$ of
the heavy neutrino we calculate which mass range will be detectable
by the upcoming CMBR experiments.
To estimate the obtainable precision we use the Fisher information
matrix \cite{tegmark}
\begin{equation}
I_{ij} = \sum_{l=2}^{l_{\rm max}} (2l+1) \left[C_l + C_{l,{\rm error}}
\right]^{-2} \frac{\partial C_l}{\partial \theta_i}
\frac{\partial C_l}{\partial \theta_j},
\end{equation}
where $C_{l,{\rm error}}$ represents the experimental error which we neglect
in the present paper.
The standard error in parameter $i$ is then
\begin{equation}
\sigma_i^2 \simeq (I^{-1})_{ii},
\end{equation}
if all cosmological parameters must be determined simultaneously, and
\begin{equation}
\sigma_i^2 \simeq (I_{ii})^{-1},
\end{equation}
if all the other parameters are assumed to have already been determined
\cite{lopez1}.
To calculate actual CMBR spectra we use the CMBFAST package designed
by Seljak and Zaldarriaga \cite{SZ96}.

To be on the conservative side we have calculated the sensitivity of the
CMBR for the case where all parameters must be determined 
simultaneously. The results are shown in Fig.~4 where the upper panel 
shows the expected accuracy of PLANCK and the lower panel that of MAP.
For very small neutrino masses the the decay cannot take place before 
$t(m/3)$. Also, the CMBR data are less sensitive to late decays. Therefore
there is a sharp and well defined minimum mass which is detectable
through CMBR measurements. This lower limit depends on the decay mode because
the energy deposited in relativistic particles depends on the specific
decay kinematics.
\begin{figure}[h]
\begin{center}
\epsfysize=10truecm\epsfbox{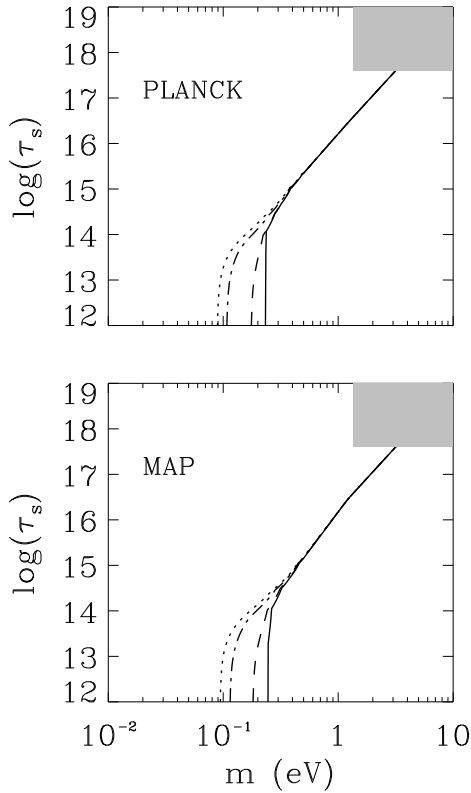}
\vspace{0truecm}
\end{center}
\vspace*{1cm}
\baselineskip 17pt
\caption{The expected detectable neutrino mass and lifetime with new
CMBR data. The region to the right of the curves is the detectable
range. Curve labels are as in Fig.~2. The shaded regions in both plots
correspond to the detectable range found by Hu, Eisenstein and Tegmark
\protect\cite{HET98}.}
\label{fig4}
\end{figure}

For both early and late decays the main effect on CMBR is through the
ISW effect \cite{tegmark}. 
This effect is due to the fact that if the universe has a
significant radiation content the linear
gravitational potential is not constant
in time. This leads to an enhancement of the CMBR fluctuations,
primarily on the scale corresponding to the horison size when the 
energy is injected.

Our results may be combined with the results of Hu, Eisenstein and Tegmark
\cite{HET98}
on stable low mass neutrinos. They estimate that for stable neutrinos
a combination of large scale structure surveys and CMBR measurements
should be able to measure a neutrino mass of 1-2 eV.

Our estimate of the obtainable precision is clearly much more restrictive 
than that found by Lopez {\it et al.} \cite{lopez2}, 
because they rely only on already existing
CMBR data, which is of low accuracy.
What we have shown is that the CMBR is a very sensitive probe of 
neutrino decays.
Clearly, our limits only apply to the case where decay takes place after
neutrinos have decoupled completely from thermal equilibrium
at $T \simeq 0.1$ MeV.


\section{Discussion}

In this paper we have calculated the constraints that it should be possible 
to put on neutrino decays using data from the new CMBR experiments.
We have paid particular attention to the different possible decay modes,
discussing both two- and three-body decays in detail. Also, the full
Boltzmann equation was solved, instead of the momentum integrated version
\cite{KT90}.
This has the clear advantage that it is possible to do calculations for
relativistic neutrino decays where inverse decays are included.

Using this formalism we estimated that it should be possible to constrain
any type of neutrino decay if the mass is of the order 1-2 eV. If the
lifetime is short it may be possible to detect neutrino masses down to
0.1 eV using this method.
Our results complement and extend existing results on CMBR constraints on
neutrino decays \cite{lopez2,hannestad2}

Finally, it is of interest to discuss how a decaying neutrino scenario
relates to the recent results from Super-Kamiokande. 
According to these
results the muon neutrino oscillates with near maximum mixing with another
neutrino, most likely the tau neutrino \cite{superK}. 
The mass difference between these
two states is of the order 0.1 eV, and if one simultaneously believes
the solar neutrino anomaly to be due to 
$\nu_e - \nu_\mu$ oscillations \cite{bahkras}
there is no room left for cosmologically interesting neutrino
decays. However, there might be other possibilities. For instance the
solar neutrino solution might be due to oscillations with a sterile neutrino
state. The only constraint is then that $\nu_\mu$ and $\nu_\tau$ should
be almost degenerate in mass. This could still allow for cosmologically
interesting and detectable decays.
Until there are more firm results on this subject, neutrino decays remain
a viable possibility.

\appendix

\section{Asymptotic behaviour}

\subsection{decay in equilibrium}

If the decay is relativistic it is a very good approximation to assume that
it proceeds in equilibrium. This means that the distribution functions
are all of equilibrium form
\begin{equation}
f_{{\rm eq},i}(E) = \frac{1}{\exp\left(\frac{E-\mu_i}{T_i}\right)+1}.
\end{equation}
In this case, instead of using the Boltzmann
equation, we can derive a very simple set of equations for the temperature
and pseudo-chemical potentials.

The procedure is the following: One starts at some initial time with given
initial conditions. Then at each timestep the distributions are evolved
forwards in time and then allowed to equilibrate completely before the next
timestep is taken.

We start with the two-body decay. In this case the equations to be solved
at each timestep are
\begin{eqnarray}
\mu_H & = & \mu_L \\
n_0 & = & n_H + n_L \\
\rho_0 & = & \rho_H + \rho_L + \rho_\phi \\
n_L & = & n_\phi. \\
T_H & = & T_L = T_\phi
\end{eqnarray}
An interesting possibility in this case is that it is possible to form
a Bose-condensate from the pseudo-scalar particle. This phenomenon has been
investigated previously in the literature \protect\cite{madsen}

For the three body decay, the corresponding equations are
\begin{eqnarray}
\mu_H & = & 3 \mu_L \\
n_0 & = & n_H + n_L/3 \\
\rho_0 & = & \rho_H + \rho_L \\
T_H & = & T_L.
\end{eqnarray}

We have solved these equations to obtain the final energy density in the
daughter products since this is the relevant quantity to know if we are
interested in effects on the CMBR.

It is of course also worth noting that for relativistic decays there is
no entropy production since all the distribution functions stay in
kinetic equilibrium throughout \cite{bernstein}. Also, the final energy
density in decay products is completely independent of the expansion
rate of the universe since equilibration is assumed to happen 
instantaneously at each temperature. In Table I we have shown the final
energy density for relativistic decays for the four different 
possibilities which we have treated.
\narrowtext
\begin{table}
\caption{Final energy density in the decay products in units of a massless
neutrino species.}

\begin{tabular}{ccdccc}
Decay & $N_{\nu_L,0}$ &$ N_{\nu, {\rm eff}}$ \\
\tableline
$\nu_H \to \nu_L \phi$ & 0 & 1.20 \\
$\nu_H \to \nu_L \phi$ & thermal & 1.48 \\
$\nu_H \to \nu_L \bar{\nu}_L \nu_L$ & 0 & 2.20 \\
$\nu_H \to \nu_L \bar{\nu}_L \nu_L$ & thermal & 2.63\\
\end{tabular}
\end{table}

\subsection{non-relativistic decays}

For strongly non-relativistic decays the behaviour is also quite simple.
Since the energy in the decaying particle is equal to the mass, $m$, we
can write an equation for the evolution of energy density in decay products
\begin{equation}
\frac{\rho}{\rho_0} = \int_{t_i}^t \frac{e^{-t/\tau}}{\tau} 
\frac{m}{\langle E_0 \rangle} dt,
\end{equation}
where $\rho_0$ is the energy density in a standard massless species.
$\langle E_0 \rangle = 3.151 \, T$ is the mean energy of a massless 
fermion and $t_i$ is some initial time which is taken to be $t_i \ll \tau$.
Solving this equation we get
\begin{equation}
\frac{\rho}{\rho_0} = \frac{270 \, \zeta(3)}{7 \, \pi^{7/2}}
 \alpha^{1/2}
\simeq 0.84 \, \alpha^{1/2},
\end{equation}
for decays in the radiation dominated era and
\begin{equation}
\frac{\rho}{\rho_0} = \frac{540 \, \zeta(3)}{7 \, \pi^4}
\Gamma\left(\frac{5}{3}\right) \alpha^{2/3}
\simeq 0.86 \, \alpha^{2/3},
\end{equation}
for decays in the matter dominated regime.


\end{document}